\documentclass[12pt,titlepage]{article}
\usepackage{graphics,latexsym}

\newcommand{\ZZ}{{\rm Z}}
\newcommand{\NN}{{\rm N}}
\newcommand{\be}{\begin{equation}}
\newcommand{\ee}{\end{equation}}

\begin{document}
\title{Cellular Automata Models for Diffusion of Innovations}
\author{Henryk Fuk\'s\thanks{Department of Physics, University of Illinois,  Chicago,
IL 60607-7059, USA}
\and Nino Boccara$^{1,}$\thanks{
DRECAM-SPEC, CE Saclay, 91191 Gif-sur-Yvette Cedex,
France}}
\maketitle
\begin{abstract}
We propose a  probabilistic cellular automata model for the  spread of
innovations,
rumors, news, etc. in a social system. The local rule used in the model
is outertotalistic, and the range of interaction can vary. When the
range $R$ of the rule increases, the takeover time for innovation
increases and
converges toward its mean-field value, which is almost inversely
proportional
to $R$ when $R$ is large. Exact solutions for $R=1$ and
$R=\infty$ (mean-field) are presented, as well as simulation results for
other values of $R$. The average local density is found to converge to a certain stationary
value, which allows us to obtain a semi-phenomenological solution
valid in the vicinity of the fixed point $n=1$ (for large $t$).

\end{abstract}

\section{Introduction}

Diffusion phenomena in social systems such as spread of news,
rumors or innovations have been extensively studied for the past
three decades by social scientists, geographers, economists, as well as
management and marketing scholars. Traditionally, ordinary differential equations
have been used to model these phenomena, beginning with the Bass model
\cite{bass69} and ending with sophisticated models which take into account
 learning, risk aversion, nature of innovation etc. \cite{Mahajan1990,
Mahajan85,Rogers95}. Models incorporating space and spatial distribution
of individuals have been also proposed, although most research in this
field has been directed to the refining of discrete Hagerstrand models
\cite{Hagerstrand52,Hagerstrand65} and to constructing partial differential
equations similar to diffusion equations known to physicists
\cite{Haynes77}.

Diffusion of innovations (we will use this term in a general sense,
meaning not only innovations, but also news, rumors, new products etc.)
is usually defined as a process by which the innovation
``is communicated through certain
channels over time among the members of a social system'' \cite{Rogers95}.
In most cases, these communication channels have rather short range,
i.e. in our decisions we are heavily influenced by our friends, family,
coworkers, but not that much by unknown people in distant cities.
This local nature of social interactions
makes cellular automata a convenient tool in modeling of diffusion phenomena.
In fact, epidemic models formulated in terms of automata networks have
been succesfully constructed in recent years \cite{bc92,bc93,bco94}.

In what follows, we will propose a model for the spread of innovations
formulated in terms of  probabilistic cellular automata, and
investigate
the role of the range of interaction in the diffusion process.

Let us define a  {\em symbol} set ${\cal G} =\{0,1,...k-1\}$ and a { bisequence}
over ${\cal G}$ which is a function on $\ZZ$ (the set of all integers)
 to ${\cal G}$, that is, $s:\ZZ \to {\cal G}$.
 The set of all bisequences over ${\cal G}$, i.e. ${\cal S}={\cal
G}^{\ZZ}$,
   is called the {\it configuration space}, and its elements are called
   {\it configurations}. The mapping
 $T: {\cal S} \to {\cal S}$
      defined by $[T(s)](i)=s(i+1)$ for every $i \in \ZZ$ is called the {\it (left) shift}.

 Let $F:{\cal S} \to {\cal S}$ be a continuous map commuting with $T$.
In general,  a {\em cellular automaton (CA)} \cite{Wolfram86}
 is a discrete dynamical
 system defined by
 \be
   s_{t+1}=F(s_t).
 \ee
As proved in \cite{hedlund69}, in this case one can always find
a {\it local mapping}  such that
the value of $s_{t+1}(i)$ depends only
on a finite number of neighboring $s_t(j)$ values.
In the remaining part of this paper we will only consider
2-state probabilistic cellular automata,
with a dynamics such that $s_{t+1}(i)$ depends on $s_t(i)$
and $\sigma_t(i)$, where
   \be
\sigma_t(i)=\sum_{n=-\infty}^{\infty}s_t(i+n)p(n),
\ee
and $p$ is a nonnegative function
 satisfying
\be
\sum_{n=-\infty}^{\infty}p(n)=1.
\ee

In the simplest version of our model, the sites of an infinite lattice are all
occupied by individuals. The individuals are of two different types:
{\em adopters} ($s_t(i)=1$) and {\em neutrals} ($s_t(i)=0$). Once an
individual becomes
an adopter, he remains an adopter, i.e. his state cannot change. At every
time
step, each neutral individual can become an adopter with a probability
depending on the parameter $\sigma$.
In this paper we will assume that this probability is equal to $\sigma$.

\begin{figure}
\begin{center}
 \scalebox{0.8}{\includegraphics{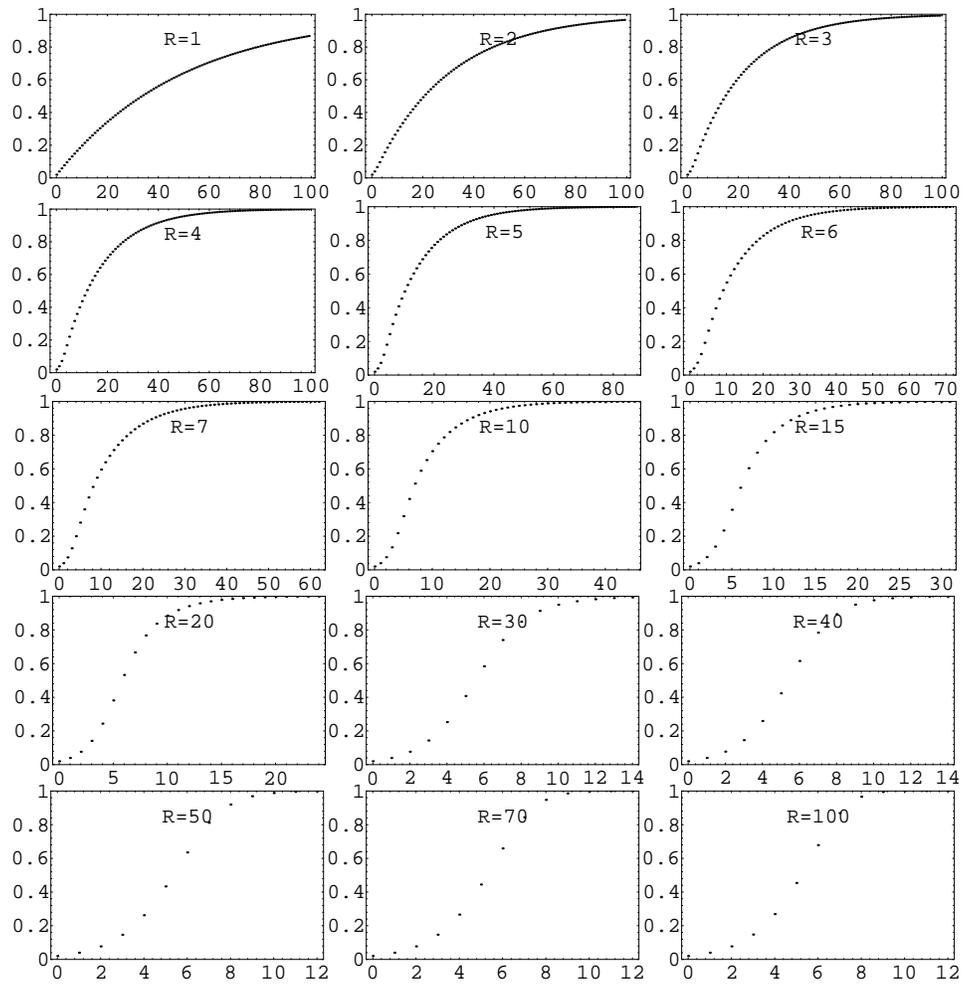}}
\end{center}
\caption{Density of adopters (vertical axis) as a function of time (horizontal
axis) for several values of
radius $R$}
\label{fignoft}
\end{figure}

The model can be viewed as a
probabilistic cellular automaton with probability distribution
\begin{eqnarray}
   P(s_{t+1}(i)=0)&=&\big(1-s_t(i)\big)\big(1-\sigma_t(i)\big) \\
   P(s_{t+1}(i)=1)&=&1-\big(1-s_t(i)\big)\big(1-\sigma_t(i)\big)
\end{eqnarray}
The {\it transition probability} $P_{b \leftarrow a}$ is defined as
\be
P_{b \leftarrow a}=P(s_{t+1}(i)=b | s_t(i)=a),
\ee
and represents the probability for a given site of changing its state from
$a$
to $b$ in one time step.
The transition probability matrix in our case has the form
\be
\mathbf{P}=
\left[
\begin{array}{cc}
 P_{0 \leftarrow 0} & P_{0 \leftarrow 1} \\
 P_{1 \leftarrow 0} & P_{1 \leftarrow 1}
\end{array}  \right] =
\left[
\begin{array}{cc}
 1-\sigma_t(i) & 0 \\
   \sigma_t(i) & 1
\end{array} \right] .
\label{probmatrix}
\ee
Subsequently, we will consider an uniform {\em outertotalistic} neighborhood of
radius~$R$, with the parameter $\sigma$ defined as
\be
\sigma_t(i)=\frac{1}{2R}\left(\sum_{n=-R}^{-1}s_t(i+n)+
\sum_{n=1}^{R}s_t(i+n)\right).
\ee
$\sigma$ is then a local density of adopters over $2R$ closest neighbors.
This choice of $\sigma$, although somewhat simplistic,  captures some
essential features of a real social system: the number of influential
neighbors
 is finite and these neighbors are all located within a certain finite
radius $R$.
Opinions of all neighbors have equal weight here, which is maybe not
realistic, but good enough as a first approximation.
Let $n_t$ be the global density of adopters at time $t$ (i.e. number of
adopters per lattice site), and $m_t=1-n_t$.
Since  $P_{1 \leftarrow 1}=1$, the number of adopters increases
with time, and $\lim_{t \rightarrow \infty} n_t=1$. If we start with a
small initial density of randomly distributed adopters $n_0$, $n_t$ follows
a characteristic S-shaped curve, typical to many growth processes.
 The curve becomes steeper when $R$ increases, and if $R$ is large enough
it takes only a few time steps to reach a high $n$ (e.g. $n=0.99$).
Figure \ref{fignoft} shows some examples of curves obtained in a computer
experiment with a
lattice size equal to $10^5$ and $n_0=0.02$.

\section{Exact Results}

Average densities $n_{t+1}$ and $m_{t+1}$ can be obtained
from previous densities $n_t$ and $m_t$ using the transition probability matrix
( $\langle \rangle$ denotes a spatial average)
\be
\left(
\begin{array}{c}
 m_{t+1}  \\
 n_{t+1}
\end{array}
\right)
=
\left[
\begin{array}{cc}
 \langle P_{0 \leftarrow 0} \rangle & \langle P_{0 \leftarrow 1} \rangle \\
 \langle P_{1 \leftarrow 0} \rangle & \langle P_{1 \leftarrow 1} \rangle
\end{array}  \right]
\left(
\begin{array}{c}
 m_{t}  \\
 n_{t}
\end{array}
\right),
\ee
hence:
\be
n_{t+1}=n_t \langle P_{1 \leftarrow 1} \rangle + (1-n_t)
\langle P_{1 \leftarrow 0} \rangle,
\ee
and using (\ref{probmatrix}) we have
\be
n_{t+1}=n_t + (1-n_t) \langle \sigma_t(i) \rangle. \label{difeq}
\ee
This difference equation can be solved in two special cases,
$R=1$ and $R=\infty$.

If $R=1$, only three possible values of local density
$\sigma$ are allowed: $0,\frac{1}{2}$ and $1$. The initial configuration
can be viewed as clusters of ones separated by clusters of
zeros. Only neutral sites adjacent to a cluster of ones can change their state,
and they will become adopters with probability $\frac{1}{2}$. All other
neutral sites have local density equal to zero, therefore they will remain
neutral, as shown in an example below (sites that can change are
underlined):
\begin{center}
$\cdots$
\begin{tabular}{|c|c|c|c|c|c|c|c|c|c|c|c|}\hline
                 1&\underline{0}&0&\underline{0}&1&1&\underline{0}&0&0&\underline{0}&1&1 \\ \hline
                \end{tabular} $\cdots$
\end{center}
This implies that the length of a cluster of zeros will on average
decrease by one every time step, i.e.
\be
M(s,t+1)=M(s+1,t),
\ee
where $M_(s,t)$ denotes a number of clusters of zeros of size $s$ at
time $t$, and furthermore
\be
M(s,t)=M(s+t,0).
\ee
For a random configuration with initial density $n_0$, the cluster density
is given by
\be
M(s,0)=(1-n_0)^s n_0^2,
\ee
hence
\be
M(s,t)=(1-n_0)^{s+t} n_0^2.
\ee
The density of zeros can be now easily computed as
\be
m_t=\sum_{s=1}^{\infty} s M(s,t) = (1-n_0)^{t+1},
\ee
and finally
\be
n_t=1-(1-n_0)^{t+1}.
\ee
Using the above expression, we obtain
\be
n_{t+1}=n_t + (1-n_t) n_0,
\ee
Comparing with (\ref{difeq}) this yields $\langle \sigma_t \rangle = n_0$, i.e.
the average probability that a neutral individual  adopts the innovation
 is time independent and equal to the initial density of adopters.

When $R \rightarrow \infty$, local density of ones becomes equivalent to
the
global density, thus in (\ref{difeq}) we can replace $\langle \sigma_t
\rangle$
 by $n_t$:
\be
n_{t+1}=n_t + (1-n_t)n_t,
\ee
or
\be
1-n_{t+1}=(1-n_t)^2.
\ee
Hence
\be
n_t=1-(1-n_0)^{2^t}.
\ee
Note that this case corresponds to the mean-field approximation, as we
neglect all spatial correlations and replace the local density by the global
one.

\section{Simulations}
\begin{figure}
\begin{center}
 \scalebox{0.8}{\includegraphics{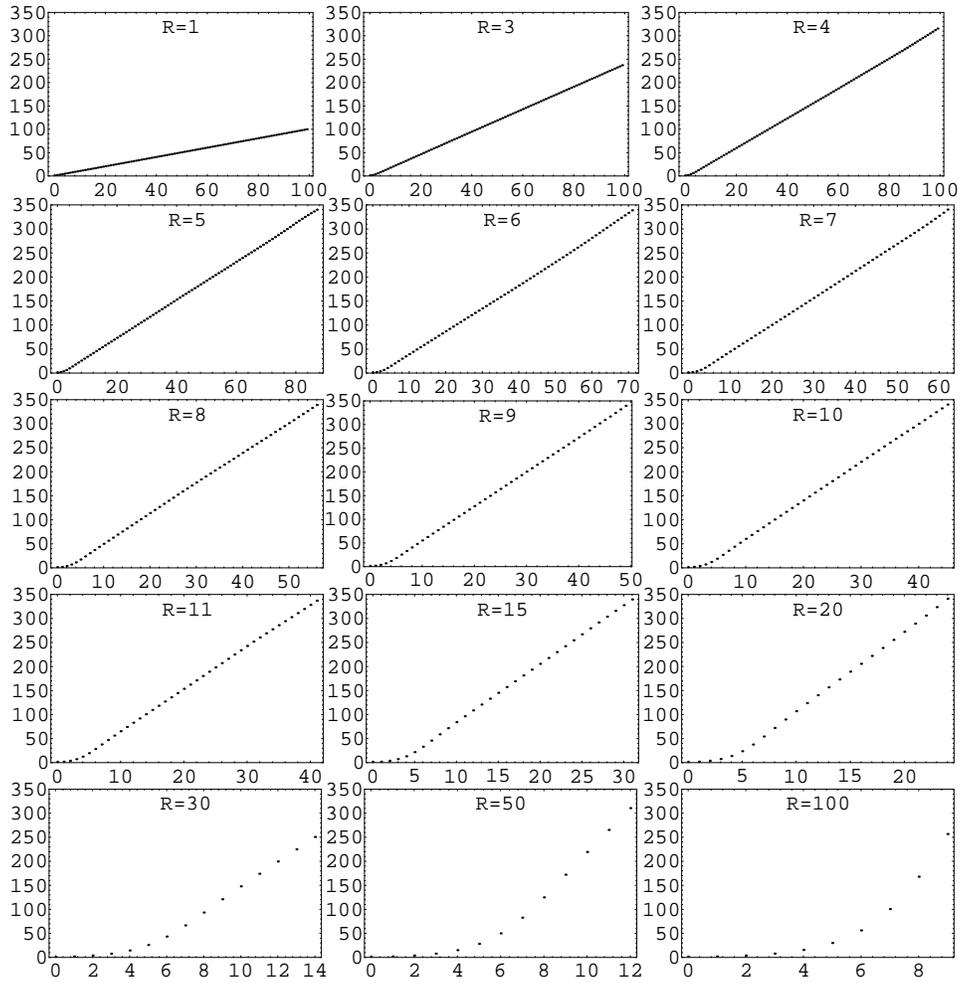}}
\end{center}
\caption{Plots of $f(t,R)$ (vertical axis) as a function of time (horizontal
axis) for several values of
radius $R$.}
\label{figfoft}
\end{figure}

Solutions obtained for the limiting cases discussed in the previous section
suggest that in general the density of adopters might have the form
\be
n_t=1-(1-n_0)^{f(t,R)}, \label{fdef}
\ee
where $f$ is a certain unknown function satisfying $f(t,1)=t+1$ and
$f(t, \infty )=2^t$. Plots of this function for several different
values of $R$ are shown in Figure~\ref{figfoft}. They were constructed by measuring
$n_t$ and  using
\be
f(t,R)=\frac{\ln (1-n_t)}{\ln (1-n_0)}.
\ee
Two apparent features of $f(t,R)$ can be observed: for every finite $R$,
the function $f(t,R)$ becomes asymptotically linear, and the slope
of the asymptote increases with $R$. This slope can be defined as a limit
\be
a(R)=\lim_{t \rightarrow \infty} \frac{\ln (1-n_t)}{t \ln (1-n_0)},
\ee
and as Figure \ref{figfoft} indicates, $a(R)$ increases with $R$.

\begin{figure}
\begin{center}
 \scalebox{0.8}{\includegraphics{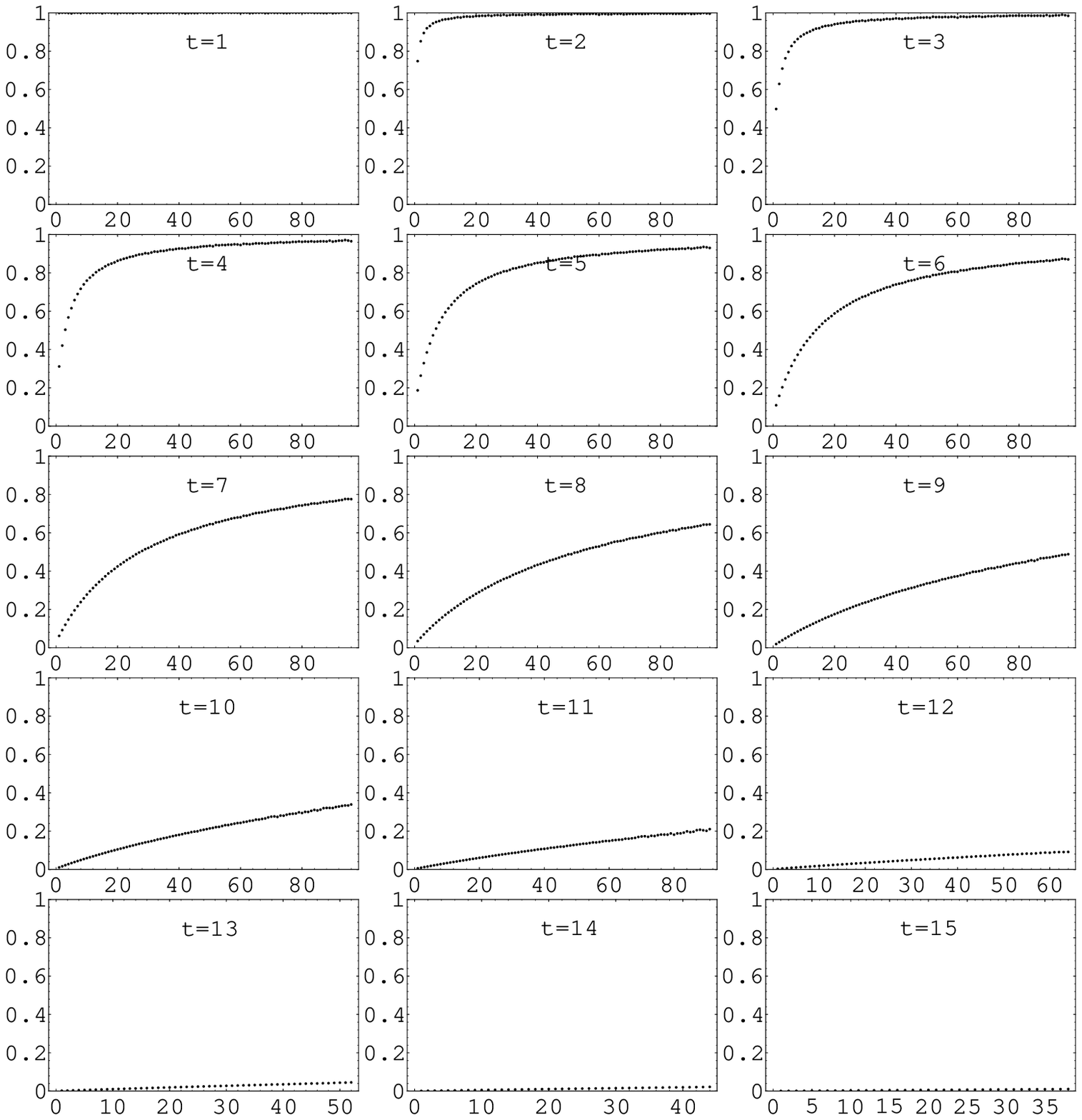}}
\end{center}
\caption{Plots of $g(t,R)=2^{-t}f(t,R)$ (vertical axis) as a function of radius $R$ (horizontal
axis) for several values of time $t$.}
\label{figgofr}
\end{figure}

Since $f(t,\infty)=2^t$, we can ``rescale'' $f(t,R)$ and introduce $g(t,R)=
2^{-t} f(t,R)$, so that $\lim_{R \rightarrow \infty} g(t,R) = 1$. Several
sample graphs of $g(t,R)$ as a function of $R$ for different values of
$t$ are shown in Figure \ref{figgofr}.
 This set of graphs reveals certain regularity
of $g(t,R)$, namely the shape of the $g(t,R)$ for a given $t$
is the same as the shape of $g(t+1,R)$ with vertical axis rescaled
by a factor 2, i.e. $g(t+1,2R) = g(t,R)$. In terms of our original
function $f$ this becomes
\be
f(t+k, 2^k R)=2^k f(t,R), \label{scaling}
\ee
where k is a positive integer.
This condition is more accurate for larger $R$, although we found it
well satisfied even for rather small $R$ (less than 10). It is seriously
violated
only for $R=1$, when $f(t,1)=t+1$.

\begin{figure}
\begin{center}
 \scalebox{0.54}{\includegraphics{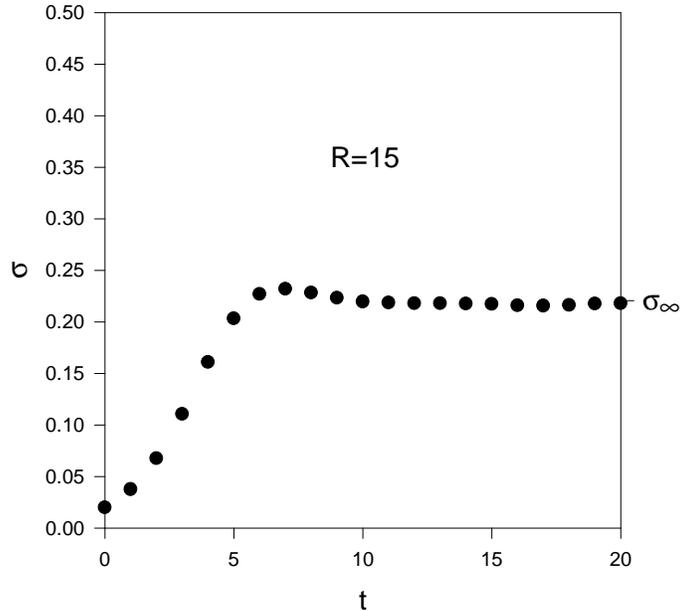}}
\end{center}
\caption{Typical plot of the $\sigma_t$ versus time, for $R=15$.}
\label{figsig15}
\end{figure}

\begin{figure}
\begin{center}
 \scalebox{0.54}{\includegraphics{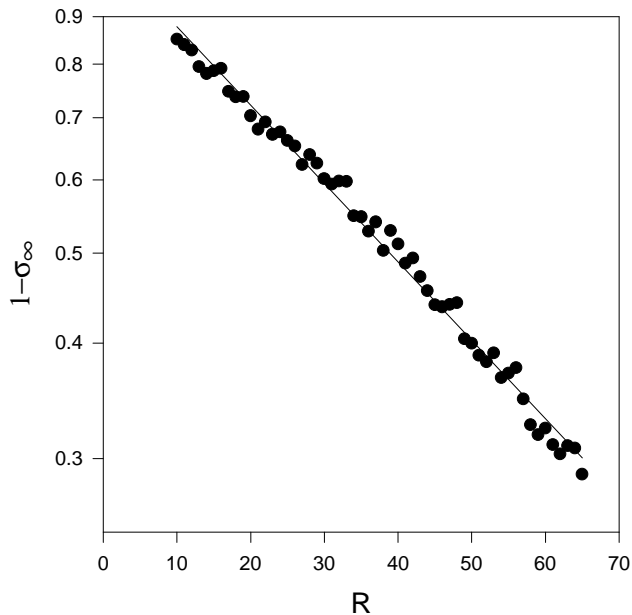}}
\end{center}
\caption{Plot of $\ln (1- \langle \sigma_{\infty} \rangle)$
as a function of $R$. }
\label{figsinf}
\end{figure}

Combining (\ref{difeq}) and (\ref{fdef}), we can express average
local density $\sigma$ in terms of $f(t,R)$:
\be
\langle \sigma_t \rangle=1-(1-n_0)^{f(t+1,R)-f(t,R)}.
\ee
Since $f(t,R)$ is linear for large $t$, the average local density must become
time-independent for large $t$, and therefore
\be
\lim_{t \rightarrow \infty} \langle \sigma_t \rangle=
1-(1-n_0)^{a(R)}. \label{sigofa}
\ee
This limit will be referred to as $\langle \sigma_{\infty} \rangle$. The average
local density converges to $\langle \sigma_{\infty} \rangle$ rather quickly.
 For example, when $R=15$, only 10 time steps are needed in practice
 to reach the limit value, as shown in Figure \ref{figsig15}.
Condition (\ref{scaling}) can be now used to find the dependence of
$\sigma_\infty$ upon the radius $R$. As we already noted, $f(t,R)$
is asymptotically linear when $t\rightarrow \infty$, and the convergence
is so fast that we can assume $f(t,R)=a(R)t + b(R)$  for large $t$.
Using condition (\ref{scaling}) we have
\be
f(t,2R)=2f(t-1,R)=2a(R)t +b(R)-a(R),
\ee
which implies that $a(2R)=2a(R)$, i.e. $a(R)$ is proportional to $R$.
This can be easily verified if we rewrite (\ref{sigofa}) as
\be
a(R)\ln(1-n_0)= \ln (1- \langle \sigma_{\infty} \rangle),
\ee
and plot the right hand side of the above equation versus $R$, as
shown in Figure \ref{figsinf}.
 Of course, the slope $a(R)$ could also be measured
directly from graphs of $f(t,R)$, and then plotted as a function of $R$.
In both cases, $a(R)=\gamma R$, where $\gamma=0.89 \pm 0.05$, and finally
\be
 \langle \sigma_\infty \rangle=
1-(1-n_0)^{\gamma R}.
\ee
As before, this expression is not very accurate if the radius $R$ is  small
(close to~1).

The value of $a(R)$, in a sense, provides us with a measure of a ``speed''
of convergence toward fixed point $n=1$, assuming we are sufficiently close
to $n=1$. It does not carry, however, useful information about the early
stage of the process, away from $n=1$ (all our previous derivations
assume large $t$).

\begin{figure}
\begin{center}
 \scalebox{0.65}{\includegraphics{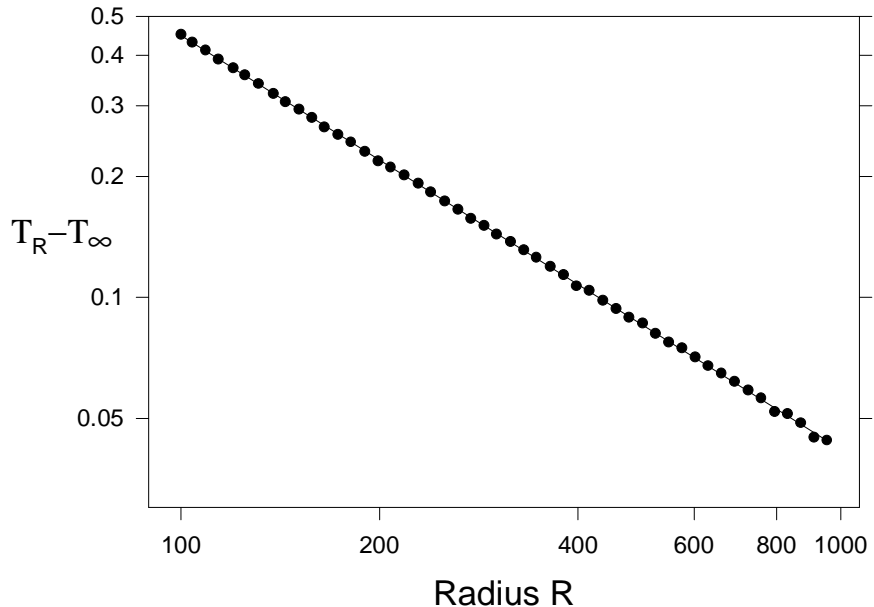}}
\end{center}
\caption{Asymptotic behavior of $T_R-T_\infty$ for large $R$.}
\label{figtotime}
\end{figure}
Therefore, we will consider another quantity, often used in marketing,
called the {\em takeover time}, and usually defined
as the time required to go from $n=0.1$ to $n=0.9$. For an S-shaped curve,
the inverse of its maximal slope carries a similar information as the
takeover time,
therefore we define
\be
T_R=\left( \max_{t \in \NN}\{(n_{t+1}-n_t)\}\right)^{-1}.
\ee
Subsequently, $T_R$ will be used as a measure of the takeover time
for a model with neighborhood radius $R$, and for simplicity
we will just call it takeover time.

For $R=\infty$, the largest $n_{t+1}-n_t$ occurs at $t=5$, therefore
\be
T_{\infty}=(n_6-n_5)^{-1}=\left((1-n_0)^{32}-(1-n_0)^{64}\right)^{-1}
\ee
We used $n_0=0.02$, which yields $T_\infty =4.00915$.

We studied the asymptotic behavior of $T_R-T_\infty$ as $R$ tends
to infinity (Figure \ref{figtotime}).
This log-log plot shows that $T_R$ tends to $T_\infty$
as $R^{-\alpha}$, where $\alpha=1.03 \pm 0.01$. This exponent depends
rather weakly on
the initial density of adopters $n_0$.
For example, if $n_0$ is two times smaller ($n_0=0.01$), $\alpha$ changes its
value by approximately $2\%$, i.e. $\alpha=1.05 \pm 0.01$. Note that in this
case $T_\infty$ is different too, namely $T_\infty=4.01051$.

\section{Conclusion and Remarks}

We have studied a probabilistic cellular automata model for the spread of
innovations. Our results emphasize the importance of the range of interaction
between individuals, as the innovation propagates faster when the
radius is larger. Increased connectivity between individuals reduces constraints
on the exchange of information, and thus the growth rate increases. It
should
be pointed out that larger connectivity can be also achieved by increasing
dimensionality of space. We performed some measurements of the takeover
time
in two and three dimensions (to be reported elsewhere), and as expected,
the takeover
time was found smaller when dimensionality increased. This is also
consistent with the fact that, in general, in higher dimensions we are closer and
closer to the mean-field approximation.

The behavior of local and global densities of adopters
is an interesting feature of the model. While the global density
of adopters always converges to $1$, the local density does not. It
quickly reaches certain ``equilibrium value'' and stays constant after that
(for infinite lattice only, since for a finite lattice both densities
reach $1$ after a finite number of steps).

More sophisticated (and realistic) models can be constructed by relaxing
some assumptions discussed in the introduction. For example, since
every technology has a finite life span, we can allow adopters to go back
to neutral state with a certain probability. Moreover, we can
incorporate ``reluctance'' to adopt by assuming that the probability of
adoption is not equal, but proportional to the local density of adopters.
Such a generalized model exhibits a second order phase transition with a
transition
point strongly dependent on $R$. These results will be reported in details
elsewhere.


\end{document}